\documentclass[11pt,twoside]{article}
\usepackage{graphicx,epsfig,natbib,epstopdf}
\usepackage{CS18}
\begin{document}
%
%
%
\title{Updating the Dartmouth Stellar Evolution Model Grid: Pre-main-sequence Models \&
	   Magnetic Fields}
%
%
\author{Gregory A. Feiden$^{1}$, Jaquille Jones$^{2}$, Brian Chaboyer$^{2}$}
\affil{$^1$Department of Physics \& Astronomy, Uppsala University, Box 516, SE-751 20, Uppsala, Sweden}
\affil{$^2$Department of Physics \& Astronomy, Dartmouth College, 6127 Wilder Laboratory, Hanover, NH 03755, USA}
\begin{abstract}
%
%
We present the current status of an effort to create an updated grid of low-mass stellar 
evolution mass tracks and isochrones computed using the Dartmouth stellar evolution code. 
Emphasis is placed on reliably extending the present grid to the pre-main-sequence, where 
modeling uncertainties have the greatest impact. Revisions to the original code release 
include: updated surface boundary conditions, the introduction of deuterium burning, and 
magnetic fields. The mass track grid contains models with a mass above 0.1 $M_{\sun}$ 
and metallicities in the range of $-0.5$ to $+0.5$~dex. Magnetic mass tracks are calculated 
for surface magnetic field strengths between 0.1~kG and 4.0~kG using two different prescriptions 
for magneto-convection. Standard and magnetic model isochrones are available for ages older 
than 1~Myr. Tabulated quantities include the stellar fundamental properties, absolute photometric 
magnitudes, magnetic field properties, convective turnover times, apsidal motion constants, 
and lithium abundances. The complete grid of mass tracks and isochrones will be made publicly 
available.
\end{abstract}
%
%
%
%
%
\section{Motivation, or ``why another grid?''}
The wealth of high precision data on stellar fundamental properties has revealed shortcomings 
in the predictive power of stellar evolution theory across the HR diagram \citep[e.g.,][Mann et al.\
this volume]{Mathieu2007,Torres2010}. Discrepancies between observations and theory
are predominantly found for low mass main-sequence (MS) and pre-main-sequence stars (PMS)
\citep[e.g.,][]{Hillenbrand2004,Ribas2006,FC12,Spada2013}, where real stars appear to have
larger radii and cooler effective temperatures ($T_{\rm eff}$). 
Interplay between magnetic fields and convection, either deep in the interior or near the stellar
photosphere, is the favored explanation \citep[e.g.,][]{MM01,
Chabrier2007}. Magnetic fields can stabilize a fluid against convection, causing a reduction
in convective flux that leads to a steeper temperature gradient and forces radiation to carry the 
excess energy. Although well established on theoretical grounds and observationally in Sun spots, 
it is not clear whether magnetic fields are causing a \emph{global} re-structuring of low mass stellar 
interiors. 

Stellar models developed to investigate the influence of magneto-convection on stellar interior
structure yield mixed results. Models of stars with radiative cores and convective outer envelopes
suggest that magnetic fields may be responsible for observed discrepancies between models and 
observations \citep{FC12b,FC13,MM14}. For models of fully convective stars, on the other hand, 
it appears that magnetic fields can influence their global structure, but only if strong interior 
magnetic field strengths are invoked \citep{MM01,MM11,MM14,FC14}. Whether the strong interior 
magnetic fields are physically plausible remains a matter of debate \citep{MM11,FC14}. However,
these studies have focused primarily on the MS, but it has been shown that PMS models are 
sensitive to magnetic field perturbations \citep{DAntona2000,Malo2014}, suggesting that PMS stars may
provide strong tests of magnetic models.

To better understand the strengths and weaknesses of these magnetic models, it is important to 
apply them in a variety of astrophysical contexts across the HR diagram to see where they succeed
and where they fail or where it is too difficult to tell. This is best achieved through the 
development of a model grid covering 
several evolutionary phases (PMS to red giant branch) with a range of magnetic field parameters 
(dynamo type, surface strength, etc) and made publicly available to the wider community. Here,
we report on progress to update the Dartmouth stellar evolution model grid\footnote{ The current 
Dartmouth model grid is at http://stellar.dartmouth.edu/models} \citep{Dotter2008} 
by extending it to the PMS phase and adding magnetic models.\footnote{ Our full poster is available 
at https://zenodo.org/record/10649}

\section{Updates to the Grid}

\begin{figure}[t]
	\centering
	\includegraphics[width=0.6\columnwidth,clip=true,trim=50mm 25mm 40mm 15mm]{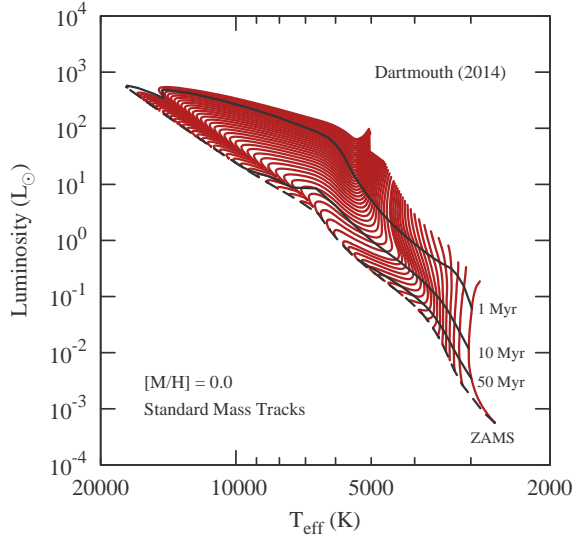}
	\caption{HR diagram for a subset of pre-main-sequence stellar evolution mass tracks at 
	solar metallicity (maroon). Masses range from 0.08 to 5.0~$M_{\sun}$. Three isochrones 
	 at 1, 10, and 50~Myr are shown as black solid lines while the zero-age-main-sequence 
	 track is a black dashed line.}
	\label{fig:massts}
\end{figure}

\subsection{Pre-Main-Sequence Models}
Before developing a grid of magnetic PMS and MS stellar models, improvements to the standard Dartmouth
stellar evolution code had to be carried out. First, deuterium burning was added by explicitly 
including the deuterium burning reaction, $^2$H($^1$H, $^3$He)$\gamma$, in the \emph{p-p} chain 
nuclear reaction network. Second, the surface boundary conditions defined by non-gray
model atmospheres had to be determined at a sufficiently large optical depth so that convection
in optically thin layers was more accurately portrayed. Though the effects on the stellar radius
are relatively insensitive to this fitting point, the stellar effective temperature derived from
the models can be influenced by up to 100 K at young ages. Instead of matching the boundary conditions
at the location where $T = T_{\rm eff}$ \citep{Dotter2007,Dotter2008}, we now adopt the location 
where the optical depth 
$\tau = 10$. Examples of mass tracks for a solar metallicity grid are shown in Figure~\ref{fig:massts}
along with representative PMS isochrones and the zero-age-main-sequence line. The full grid, 
still under development, will include masses between 0.08 -- 5.0~$M_{\sun}$ with a mass resolution 
of 0.02~$M_{\sun}$ up to 1.2~$M_{\sun}$ at 11 metallicities between $-0.5$ and +0.5~dex. Isochrone 
ages will range from 1~Myr up to 250~Myr (where the present Dartmouth grid begins) with an age 
resolution of 0.1~Myr between 1 and 20~Myr.

\subsection{Magnetic Models}
Magnetic models will be computed for a more limited set of masses between 0.1 and 1.2~$M_{\sun}$
with an initial resolution of 0.05~$M_{\sun}$. The initial release of the grid will have a 
smaller range of metallicities clustered around the solar value. However, this will be expanded
after the release. Models implementing the two ``dynamo mechanisms'' (rotational vs. turbulent)
described in \citet{FC12b,FC13} will be computed with different grids of surface magnetic field 
strengths. For the turbulent dynamo prescription, we will create models with surface magnetic 
field strengths of 0.1 to 0.9~kG  with a spacing of 0.2~kG, and for the rotational dynamo 
prescription, surface magnetic field strengths of 1.0 to 4.0~kG will be made with a spacing of 
0.5~kG. Both PMS and MS magnetic models will be made available. Presently, we have completed the 
full set of magnetic models at solar metallicity.

\section{Data Products}

\subsection{Mass Tracks}
Once completed, the full set of stellar evolution mass tracks will be made publicly available.
Mass tracks will provide evolutionary information regarding the stellar fundamental properties 
(radius, $T_{\rm eff}$, luminosity, $\log g$), properties of the stellar convection zone(s) (mass,
radial boundary, overturn timescales), the apsidal motion constant $k_2$, and surface abundances 
of the light elements deuterium and lithium. Photometric magnitudes, if not provided, will be made 
accessible via web-based post-processing routines.

\begin{figure}[t]
	\centering
	\includegraphics[width=0.65\columnwidth,clip=true,trim=10mm 30mm 40mm 90mm]{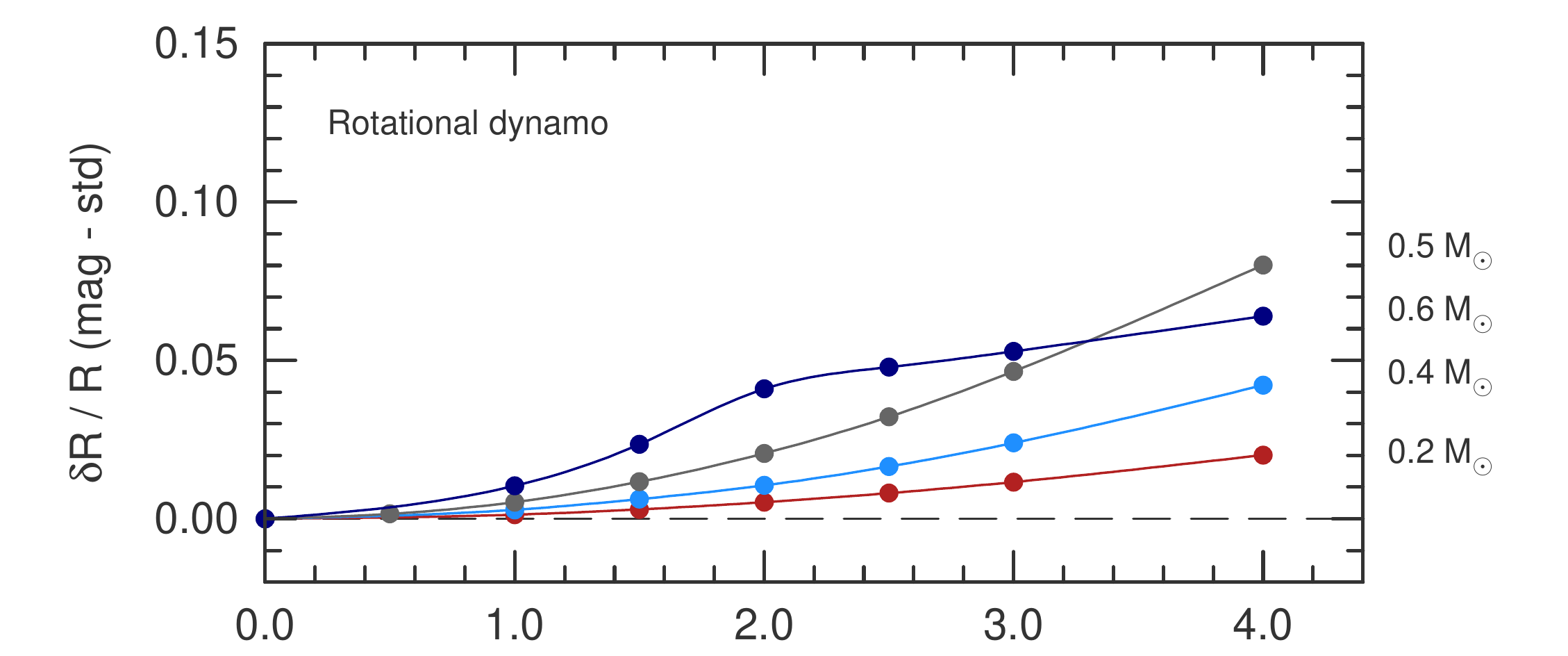} \\
	\includegraphics[width=0.65\columnwidth,clip=true,trim=10mm 20mm 40mm 90mm]{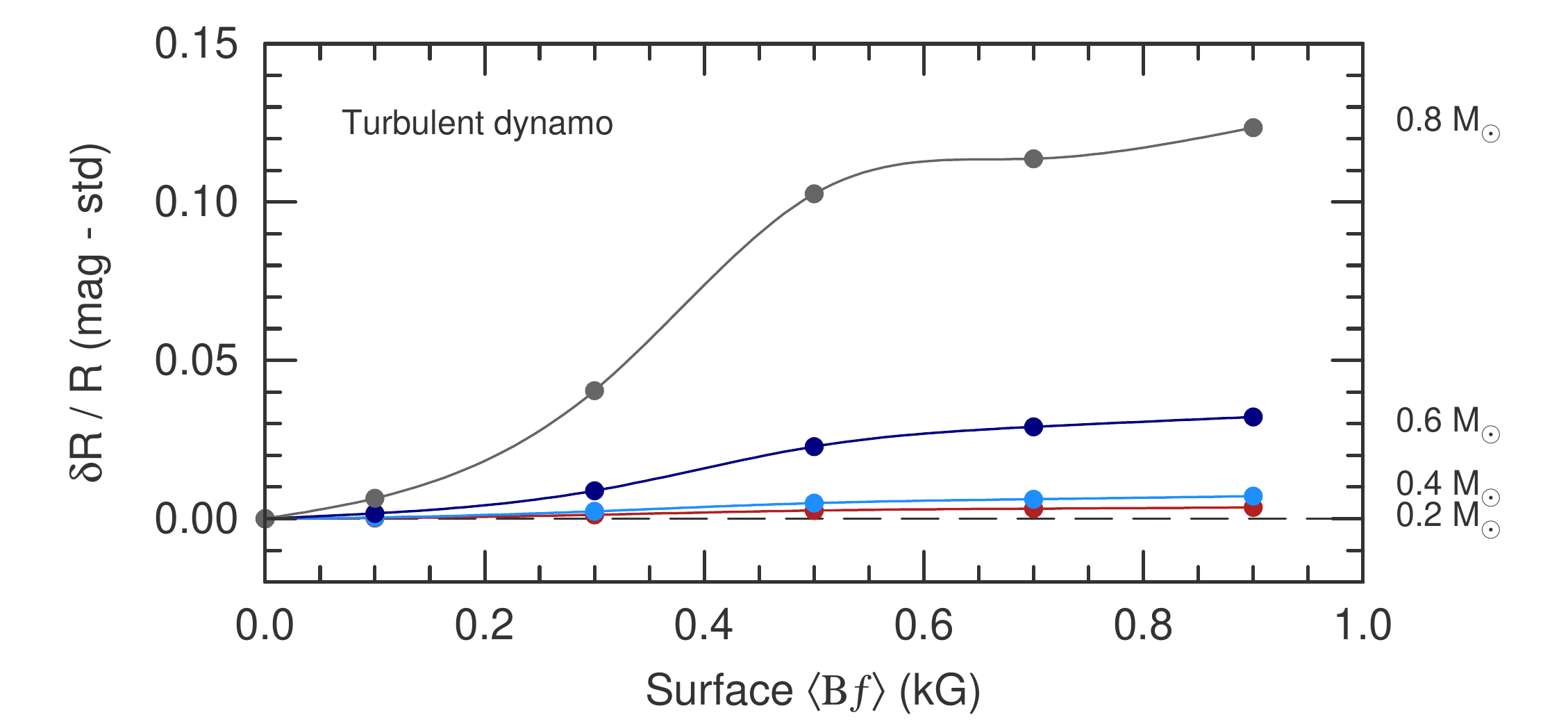}
	\caption{Relative difference (inflation) between magnetic and standard stellar evolution models
	at an age of 1~Gyr with solar metallicity. Masses are labelled to the right of their associated
	inflation curve. (\emph{top}) Models computed with a rotational dynamo prescription. (\emph{bottom})
	Models computed with a turbulent dynamo formulation. }
	\label{fig:radinf}
\end{figure}

\subsection{Inflation \& Cooling Relations}
Aside from mass tracks, a set of relations will be published that provide an estimate of radius 
inflation and temperature suppression expected for a model of a given mass and age for a 
specified surface magnetic field strength. Examples of radius inflation curves are shown 
in Figure~\ref{fig:radinf} at a few different masses using both dynamo types.

\acknowledgments{
We thank the organizers, the LOC, and the SOC for a successful Cool Stars 18 and for giving us 
the opportunity to present our work.
The Dartmouth stellar evolution model grid was originally developed by A.\ Dotter and B.C. with 
support from the NSF grant AST-0094231. Subsequent developments of the magnetic Dartmouth code
were supported by the NSF grant AST-0908345 and the William H.\ Neukom 1964 Institute for 
Computational Science at Dartmouth College. This research made use of NASA's Astrophysics Data 
System. G.A.F thanks the Gnuplot development team for their continued work on the Gnuplot graphing
utility and A.\ Irwin for developing and distributing the FreeEOS.
}

\normalsize


%
%
%
%
%
\bibliographystyle{apj}

\begin{thebibliography}{17}
\expandafter\ifx\csname natexlab\endcsname\relax\def\natexlab#1{#1}\fi

%
\bibitem[{Chabrier {et~al.}(2007)Chabrier, Gallardo, \& Baraffe}]{Chabrier2007}
Chabrier, G., Gallardo, J., \& Baraffe, I. 2007, \aap, 472, L17

\bibitem[{{D'Antona} {et~al.}(2000){D'Antona}, {Ventura}, \&
  {Mazzitelli}}]{DAntona2000}
{D'Antona}, F., {Ventura}, P., \& {Mazzitelli}, I. 2000, \apj, 543, L77

\bibitem[{Dotter {et~al.}(2007)Dotter, Chaboyer, Jevremovi\'{c}, Baron,
  Ferguson, Sarajedini, \& Anderson}]{Dotter2007}
Dotter, A., Chaboyer, B., Jevremovi\'{c}, D., Baron, E., Ferguson, J.~W.,
  Sarajedini, A., \& Anderson, J. 2007, \aj, 134, 376

\bibitem[{Dotter {et~al.}(2008)Dotter, Chaboyer, Jevremovi\'{c}, Kostov, Baron,
  \& Ferguson}]{Dotter2008}
Dotter, A., Chaboyer, B., Jevremovi\'{c}, D., Kostov, V., Baron, E., \&
  Ferguson, J.~W. 2008, \apjs, 178, 89

\bibitem[{{Feiden} \& {Chaboyer}(2012{\natexlab{a}})}]{FC12}
{Feiden}, G.~A., \& {Chaboyer}, B. 2012{\natexlab{a}}, \apj, 757, 42

\bibitem[{{Feiden} \& {Chaboyer}(2012{\natexlab{b}})}]{FC12b}
{Feiden}, G.~A., \& {Chaboyer}, B. 2012{\natexlab{b}}, \apj, 761, 30

\bibitem[{{Feiden} \& {Chaboyer}(2013)}]{FC13}
{Feiden}, G.~A., \& {Chaboyer}, B. 2013, \apj, 779, 183

\bibitem[{{Feiden} \& {Chaboyer}(2014)}]{FC14}
{Feiden}, G.~A., \& {Chaboyer}, B. 2014, \apj, 786, 53

\bibitem[{{Hillenbrand} \& {White}(2004)}]{Hillenbrand2004}
{Hillenbrand}, L.~A., \& {White}, R.~J. 2004, \apj, 604, 741

\bibitem[{{MacDonald} \& {Mullan}(2012)}]{MM11}
{MacDonald}, J., \& {Mullan}, D.~J. 2012, \mnras, 421, 3084

\bibitem[{{MacDonald} \& {Mullan}(2014)}]{MM14}
{MacDonald}, J., \& {Mullan}, D.~J. 2014, \apj, 787, 70

\bibitem[{{Malo} {et~al.}(2014){Malo}, {Doyon}, {Feiden}, {Albert},
  {Lafreni{\`e}re}, {Artigau}, {Gagn{\'e}}, \& {Riedel}}]{Malo2014}
{Malo}, L., {Doyon}, R., {Feiden}, G.~A., {Albert}, L., {Lafreni{\`e}re}, D.,
  {Artigau}, {\'E}., {Gagn{\'e}}, J., \& {Riedel}, A. 2014, arXiv: 1406.6750 

\bibitem[{Mathieu {et~al.}(2007)Mathieu, Baraffe, Simon, Stassun, \&
  White}]{Mathieu2007}
Mathieu, R.~D., Baraffe, I., Simon, M., Stassun, K.~G., \& White, R. 2007, in
  Protostars \& Planets V, 411--425

\bibitem[{Mullan \& MacDonald(2001)}]{MM01}
Mullan, D.~J., \& MacDonald, J. 2001, \apj, 559, 353

\bibitem[{{Ribas}(2006)}]{Ribas2006}
{Ribas}, I. 2006, \apss, 304, 89

\bibitem[{{Spada} {et~al.}(2013){Spada}, {Demarque}, {Kim}, \&
  {Sills}}]{Spada2013}
{Spada}, F., {Demarque}, P., {Kim}, Y.-C., \& {Sills}, A. 2013, \apj, 776, 87

\bibitem[{Torres {et~al.}(2010)Torres, Andersen, \& Gim\'{e}nez}]{Torres2010}
Torres, G., Andersen, J., \& Gim\'{e}nez, A. 2010, \aapr, 18, 67
  
%
\end{thebibliography}


\end{document}